\documentclass[twocolumn,floatfix,aps,eqsecnum,superscriptaddress,prb]{revtex4-2}
\usepackage{graphicx}
\usepackage{amsmath}
\usepackage{amssymb}
\usepackage{bm}
\usepackage{hyperref}

\usepackage{color}

\begin{document}

\title{Monitored quantum transport through a disordered one-dimensional conductor}
\author{J. S\'{a}nchez Fern\'{a}n}
\affiliation{Instituut-Lorentz, Universiteit Leiden, P.O. Box 9506, 2300 RA Leiden, The Netherlands}
\author{J. Tworzyd{\l}o}
\affiliation{Faculty of Physics, University of Warsaw, ul.\ Pasteura 5, 02--093 Warszawa, Poland}
\author{C. W. J. Beenakker}
\affiliation{Instituut-Lorentz, Universiteit Leiden, P.O. Box 9506, 2300 RA Leiden, The Netherlands}

\date{May 2026}

\begin{abstract}
We formulate a quantum master equation for the many-particle density matrix of electrons propagating through a single-mode conductor, combining elastic scattering by disorder with time-resolved projective measurements that monitor the outcome of scattering events. The full counting statistics of transmitted electrons has a binomial distribution function, whose mean ${\cal T}$ and variance ${\cal T}(1-{\cal T})$ determine the conductance and shot noise power, respectively. Monitoring suppresses the phase coherence responsible for one-dimensional localization: The decay with conductor length $L$ of the typical transmission probability crosses over at $L\simeq \ell_\phi$ from the exponential $e^{-L/\xi}$ (with localization length $\xi$) to the Ohmic $1/L$ decay. Numerical solution of the master equation gives, for weak monitoring, a logarithmic dependence $\ell_\phi\simeq \xi\ln(v_{\rm F}\tau_\phi/\xi)$ of the coherence length $\ell_\phi$ on the mean time $\tau_\phi$ between measurements.
\end{abstract}
\maketitle

\section{Introduction}

The statistics of wave localization in a disordered one-dimensional (1D) system was studied in the 1950s in the context of radio engineering, and a full solution for weak backscattering was obtained by Gertsenshtein and Vasil'ev \cite{Ger59} (later reproduced in the electronic context by Abrikosov \cite{Abr81}). Unlike in the 3D case studied by Anderson \cite{And58}, in 1D the transmission probability ${\cal T}$ decays exponentially with system length $L$ for arbitrarily weak disorder.

The wave interference that is at the origin of the localization requires phase coherence. A coupling to the environment introduces a finite phase coherence length $\ell_\phi$, and the exponential decay $\propto e^{-L/\xi}$ is expected to slow down to the Ohmic $1/L$ decay characteristic of classical diffusion for $L\gg\ell_\phi$. This transition from localization to diffusion has been studied extensively \cite{DAm90,Pas91,Mas94,Kni99,Li02,Roy07a,Roy07b,Cat10,Ste14}  in the framework of B\"{u}ttiker's  voltage probe model of dephasing \cite{But86,But88}. This is a phenomenological model, with microscopic justification \cite{McL91,Her91,Gol07,Maa09}, where fictitious leads (``voltage probes'') couple the system to reservoirs in thermal equilibrium at a voltage adjusted such that each lead draws no current. The lead thus absorbs and re-emits carriers with random phase, modeling the coupling to an environment without specifying its microscopic degrees of freedom. Similar results are obtained in models that introduce random phase fluctuations in the scattering events \cite{Pal04,Zhe06,Ray13}.

These approaches are formulated in terms of single-particle wave functions, which gives access to expectation values of bilinears in the fermion operators. This is sufficient for the time-averaged current and hence the conductance. The shot noise power needs the second moment of the transferred charge --- a product of four fermion operators \cite{Bla00}. Higher moments correspond to higher moments of the fermion operators, which require knowledge of the many-particle density matrix. Here we formulate a phenomenological model for decoherence in a disordered system that addresses the full density matrix. When applied to the conductance, it gives similar results for the localization-to-diffusion crossover as the voltage probe model. In addition, it allows to calculate the full counting statistics of the transferred charge \cite{Bel04}.

Our approach is in line with recent work on monitored quantum transport \cite{Jin22,Szy23,Fer24,Tho24,Pic25,Gur25,Bee25,Cha26,Gon26}, in which a quantum master equation describes how unitary dynamics is interrupted by projective measurements. In the next section we generalize the quantum channel formulation of Ref.\ \onlinecite{Bee25}, which assumed unidirectional (chiral) transport, to include backscattering so that it can describe localization. The probability distribution function of the transferred charge is computed in Sec.\ \ref{sec_chargetransfer}. The measurements do not change the binomial form of the distribution \cite{Lev96}, they only modify the transmission probability ${\cal T}$. In particular, the shot noise power remains $\propto {\cal T}(1-{\cal T})$ in the presence of decoherence. 

A quantum master equation for ${\cal T}$ is derived in Sec.\ \ref{sec_transmission} and formally solved in Sec.\ \ref{sec_solution}. A key step here is the algebraic resummation of the exponentially many measurement outcomes that label the quantum trajectories. Applications follow in Secs.\ \ref{sec_tunnel} (to resonant tunneling, which can be evaluated analytically) and \ref{sec_localization} (to localization, which needs a numerical evaluation). We conclude in Sec.\ \ref{sec_conclude}.
 
\section{Quantum channel formulation}
\label{sec_quantumchannel}

\subsection{Scattering operators}

\begin{figure}[tb]
\centerline{\includegraphics[width=0.6\linewidth]{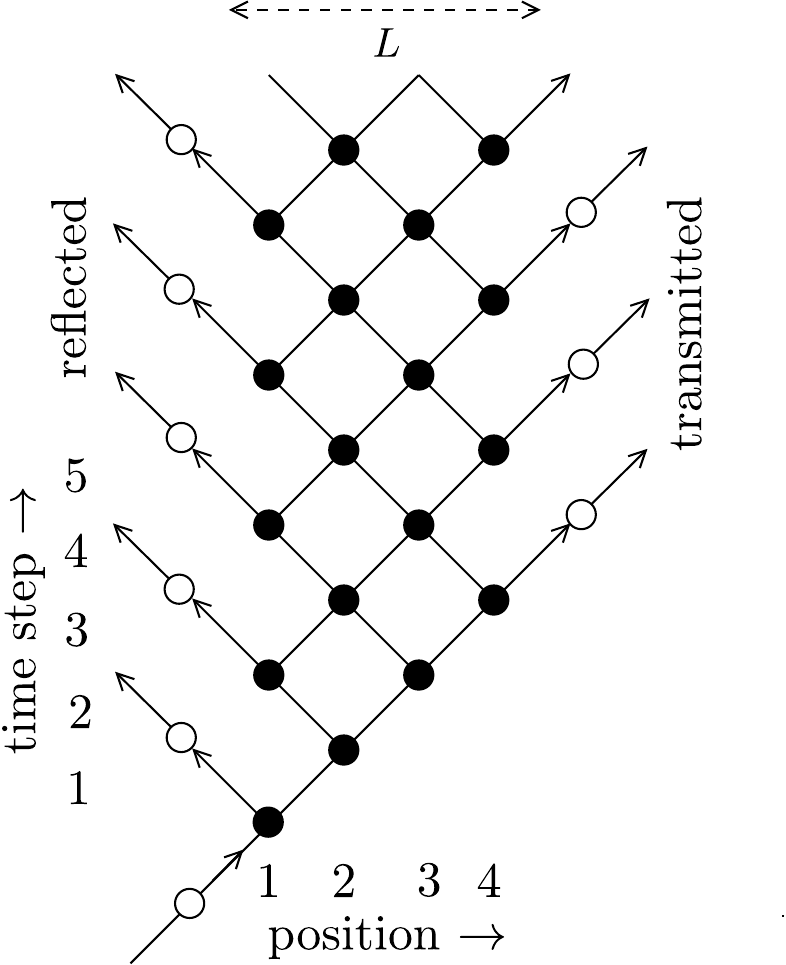}}
\caption{Representation on a 1+1 dimensional space-time lattice of transport through a single-mode disordered conductor. Scattering centra (extending over a length $L$) are indicated by filled circles, while open circles represent the ideal leads (no backscattering). Electrons enter the conductor from one end and are either reflected back or transmitted to the other end.
}
\label{fig_diagram}
\end{figure}

We represent a single-mode electronic conductor by a quantum channel, a completely-positive trace-preserving map that relates the density matrix $\rho$ of incoming electrons to that of outgoing electrons. For that purpose it is convenient to discretize space and time on a 1+1 dimensional lattice (unit lattice constant, see Fig.\ \ref{fig_diagram}). The set of fermion creation and annihilation operators is $a_{n\sigma}^\dagger,a_{n\sigma}^{\vphantom{\dagger}}$, where $n\in\mathbb{Z}$ labels the lattice site and the pseudospin $\sigma=\pm$ identifies the direction of motion along the lattice (left-mover or right-mover). The real electron spin plays no role and is ignored.

Scattering events that couple left-movers and right-movers are described by a unitary operator $\hat{U}_n$, different on different sites but not time-dependent. This operator in Fock space has the form
\begin{equation}
\hat{U}_n=\exp\left(i\sum_{\sigma,\sigma'}a^\dagger_{n\sigma}M^{(n)}_{\sigma\sigma'}a_{n\sigma'}\right),\label{hatUdef}
\end{equation}
with a $2\times 2$ Hermitian matrix $M^{(n)}$, corresponding to the single-particle scattering matrix $U_n=e^{iM^{(n)}}\in{\rm U}(2)$. 

At each time step left-movers and right-movers are displaced by one lattice site via the unitary translation operator
\begin{equation}
\hat{T}=\exp\left(i\sum_{k,\sigma} \sigma ka^\dagger_\sigma(k)a^{\vphantom{\dagger}}_\sigma(k)\right),\;\;a_\sigma(k)=\sum_n e^{-ikn}a_{n\sigma}.
\end{equation}
The scattering centra are present in a segment of length $L$, containing the sites $n=1,2,\ldots L$. The sites with $n<1$ and $n>L$ represent ideal leads, without backscattering; we set $\hat{U}_n$ equal to the identity on those sites.

The phase coherent dynamics in one time step is represented by unitary conjugation,
\begin{equation}
{\rho}(t+1)=\bigl(\hat{T} \prod_{n}\hat{U}_n\bigr)\rho(t)\bigl(\hat{T} \prod_{n}\hat{U}_n\bigr)^\dagger.\label{rhotplus1coherent}
\end{equation}
Phase coherence is lost by coupling to the environment, which effectively ``monitors'' the dynamics by measuring the outcome of each scattering event: is the electron scattered into a left-mover or a right-mover?

\subsection{Measurement operators}

Measurements replace the mapping \eqref{rhotplus1coherent} by a sum over different measurement outcomes,
\begin{equation}
{\rho}(t+1)=\sum_{\bm s}\hat{K}_{\bm s}{\rho}(t)\hat{K}^\dagger_{\bm s},\label{rhotplus1incoherent}
\end{equation}
labeled by the string $\bm{s}=(s_1,s_2,\ldots s_L)$. Iteration of this one-step evolution gives
\begin{equation}
\rho(t)=\sum_{{\bm s}_1,{\bm s}_2\ldots{\bm s}_t}\bigl(\hat{K}_{{\bm s}_t}\cdots\hat{K}_{{\bm s}_2}\hat{K}_{{\bm s}_1}\bigr) {\rho}(0)\bigl(\hat{K}_{{\bm s}_1}^\dagger\hat{K}_{{\bm s}_2}^\dagger\cdots\hat{K}_{{\bm s}_t}^\dagger\bigr).\label{rhotrho0}
\end{equation}
An individual term in the sum is referred to as a ``quantum trajectory''.

The Kraus operator $\hat{K}_{\bm s}$ has the form
\begin{equation}
\begin{split}
&\hat{K}_{\bm s}= \hat{T}\prod_{n}\hat{P}_{s_n,n}\hat{U}_n,\\
&\sum_{s_n} \hat{P}_{s_n,n}^\dagger \hat{P}^{\vphantom{\dagger}}_{s_n,n}=\hat{I}\Rightarrow\sum_{\bm s}\hat{K}^\dagger_{\bm s}\hat{K}^{\vphantom{\dagger}}_{\bm s}=\hat{I}.
\end{split}\label{Krausdef}
\end{equation}
The sum rule for the measurement operators $\hat{P}$ defines a ``positive operator-valued measure'' (POVM), the resulting sum rule for the Kraus operators $\hat{K}$ ensures that the mapping \eqref{rhotplus1incoherent} is trace preserving.

We have four measurement outcomes $s_n\in\{++,--,+-,-+\}$ per site, corresponding to a projection onto a filled $(+)$ or empty ($-$) right-moving state (R), followed by a projection onto a filled or empty left-moving state (L). To be able to tune the decoherence strength \cite{note1}, we consider a ``weak'' measurement on site $n$, interpolating with weight ${w}_n\in(0,1)$ between the identity and a projection. We will later set ${w}_n=0$ in the ideal leads, and take a constant ${w}_n\equiv {w}$ in the scattering region $1\leq n\leq L$ (the same strength for left-movers and right-movers).

The four weak measurement operators $\hat{P}_{s,s',n}=\hat{P}^{\rm L}_{s,n}\hat{P}^{\rm R}_{s',n}$ on site $n$ (with $s,s'\in\{+,-\}$) are given by products of
\begin{subequations}
\label{PRLdef}
\begin{align}
&\hat{P}^{\rm R}_{+,n}=\beta_n\hat{I} +(\alpha_n-\beta_n) a_{n+}^\dagger a_{n+}^{\vphantom{\dagger}},\\
&\hat{P}^{\rm R}_{-,n}=\beta_n\hat{I}+(\alpha_n-\beta_n) a_{n+}^{\vphantom{\dagger}}a_{n+}^\dagger,\\
&\hat{P}^{\rm L}_{+,n}=\beta_n\hat{I} +(\alpha_n-\beta_n) a_{n-}^\dagger a_{n-}^{\vphantom{\dagger}},\\
&\hat{P}^{\rm L}_{-,n}=\beta_n\hat{I}+(\alpha_n-\beta_n) a_{n-}^{\vphantom{\dagger}}a_{n-}^\dagger,\\
&\alpha_n=2^{-1/2}\sqrt{1+{w}_n},\;\;\beta_n=2^{-1/2}\sqrt{1-{w}_n}.
\end{align}
\end{subequations}
The corresponding POVM for right-moving states has operators
\begin{equation}
\begin{split}
&\hat{E}^{\rm R}_{+, n}=(\hat{P}^{\rm R}_{+,n})^\dagger \hat{P}^{\rm R}_{+,n}=\tfrac{1}{2}(1-{w}_n)\hat{I}+{w}_n a_{n+}^\dagger a_{n+}^{\vphantom{\dagger}},\\
&\hat{E}^{\rm R}_{-,n }=(\hat{P}^{\rm R}_{-,n})^\dagger \hat{P}^{\rm R}_{-,n}=\tfrac{1}{2}(1-{w}_n)\hat{I}+{w}_n a_{n+}^{\vphantom{\dagger}}a_{n+}^\dagger ,
\end{split}
\end{equation}
summing to the identity (and similarly for left-moving states).

\section{Charge transfer statistics}
\label{sec_chargetransfer}

A voltage bias $V$ injects electrons into the left lead, in an energy window $eV$ above the Fermi level $E_{\rm F}$. At zero temperature all charges in this energy window in the right lead are right-movers originating from the left lead, counted by the operator
\begin{equation}
\hat{Q}=\sum_{n=L+1}^\infty a^\dagger_{n+}a_{n+}^{\vphantom{\dagger}}.
\end{equation}

Because the electrons are assumed to be noninteracting, it is sufficient to consider first a single injected charge, which at $t=0$ is spread out in the left lead over a length $L_0\simeq \hbar v_{\rm F}/eV$. We work in the low-voltage, linear response regime $eV\ll  \hbar/t_{\rm dwell}$, with $t_{\rm dwell}$ the characteristic dwell time of an electron in the scattering region. In the zero-temperature limit the injected charge is in a pure state,
\begin{equation}
|\Psi_0\rangle=(1+L_0)^{-1/2}\sum_{n=-L_0}^0|n,+\rangle,\label{Psi0def}
\end{equation}
with density matrix
\begin{equation}
\rho(0)=(1+L_0)^{-1}\sum_{n,m=-L_0}^0 a_{n+}^{\dagger}|\emptyset\rangle\langle\emptyset| a_{m+}^{\vphantom{\dagger}}.
\end{equation}
Here $|\emptyset\rangle\langle\emptyset|$ is the projector onto vacuum (the unperturbed Fermi sea).

The moment generating function of the transferred charge is
\begin{equation}
F(\xi)=\lim_{t\rightarrow\infty}\operatorname{Tr}e^{\xi\hat{Q}}\rho(t).
\end{equation}
We substitute
\begin{equation}
e^{\xi \hat{Q}}=\prod_{n=L+1}^\infty\bigl[1+(e^\xi-1)a^\dagger_{n+}a_{n+}^{\vphantom{\dagger}}\bigr],
\end{equation}
which gives
\begin{equation}
\begin{split}
&F(\xi)=1+(e^\xi-1){\cal T},\\
&{\cal T}=\lim_{t\rightarrow\infty}\sum_{n=L+1}^\infty \operatorname{Tr}a^\dagger_{n+}a_{n+}^{\vphantom{\dagger}}\rho(t),
\end{split}\label{Fxibinomial}
\end{equation}
since $a_{n+}a_{m+}\rho(t)=0$ for any two sites $n,m$ (the density matrix is fully in the single-particle sector of Fock space).

Eq.\ \eqref{Fxibinomial} expresses binomial statistics of the transferred charge for a single injected particle, with transmission probability ${\cal T}$. The total number of injected particles in a long time $t_{\rm counting}$ is ${\cal N}_V=eVt_{\rm counting}/h\gg 1$, which gives the binomial probability density function \cite{Lev96}
\begin{equation}
P(Q)=\binom{{\cal N}_{V}}{Q}{\cal T}^Q(1-{\cal T})^{{\cal N}_{V}-Q}.\label{binomial}
\end{equation}
The mean and variance are
\begin{equation}
\mathbb{E}[Q]={\cal N}_V{\cal T},\;\;\operatorname{Var}[Q]={\cal N}_V{\cal T}(1-{\cal T}),
\end{equation}
corresponding to the conductance and shot noise power
\begin{equation}
G=G_0{\cal T},\;\;P_{\rm noise}=P_0{\cal T}(1-{\cal T}),\label{GPdef}
\end{equation}
with $P_0=eVG_0$ and $G_0=2e^2/h$ the conductance quantum (including spin degeneracy). 

\section{Transmission probability}
\label{sec_transmission}

The full probability distribution \eqref{binomial} of transferred charge is determined by the transmission probability ${\cal T}$. Since ${\cal T}$ is determined by the single-particle sector of Fock space, we can work with matrices in site-pseudospin vector space (first quantization), rather than with creation and annihilation operators in Fock space (second quantization).

\subsection{From Fock space to vector space}

The Kraus operator \eqref{Krausdef} in Fock space is mapped to the Kraus matrix ${\cal K}_{\bm s}$ in vector space, acting on $L$ site and 2 pseudospin variables. The unitary operators $\hat{U}_n$ map to $2L\times 2L$ unitary matrices
\begin{equation}
{\cal U}_n= |n\rangle\langle n|\otimes U_n+(1-|n\rangle\langle n|)\otimes \sigma_0.
\end{equation}
Here $|n\rangle\langle n|$ projects onto site $n$, the matrix $U_n\in {\rm U}(2)$ acts on the pseudospin (coupling left-movers and right-movers), and $\sigma_0$ denotes the $2\times 2$ unit matrix. The translation operator $\hat{T}$ maps to the shift matrix $e^{i\hat{k}\sigma_z}$ defined by
\begin{equation}
e^{i\hat{k}\sigma_z}|n,\sigma\rangle=|n+\sigma,\sigma\rangle.
\end{equation} 

The projection operators $\hat{P}_{\pm,n}^{\rm R}$ onto right-moving states map to 
\begin{equation}
\begin{split}
&{\cal P}^{\rm R}_{+,n}= |n\rangle\langle n|\otimes\begin{pmatrix}
\alpha_n&0\\
0&\beta_n
\end{pmatrix}+\beta_n(1-|n\rangle\langle n|)\otimes\sigma_0,\\
&{\cal P}^{\rm R}_{-,n}= |n\rangle\langle n|\otimes\begin{pmatrix}
\beta_n&0\\
0&\alpha_n
\end{pmatrix}+\alpha_n(1-|n\rangle\langle n|)\otimes\sigma_0.
\end{split}
\end{equation}
The projection operators onto left-moving states map to
\begin{equation}
{\cal P}^{\rm L}_{+,n}= \sigma_x{\cal P}^{\rm R}_{+,n} \sigma_x,\;\;{\cal P}^{\rm L}_{-,n}= \sigma_x {\cal P}^{\rm R}_{-,n}\sigma_x.
\end{equation}

Combining these expressions, the Kraus matrices are given by
\begin{align}
&{\cal K}_{\bm s}=e^{i\hat{k}\sigma_z}\prod_{n} {\cal A}_{s_n,n},\;\;s_n\in\{1,2,3,4\},\\
\begin{split}
&{\cal A}_{1,n}={\cal P}^{\rm L}_{+,n}{\cal P}^{\rm R}_{+,n}{\cal U}_n,\;\;
{\cal A}_{2,n}={\cal P}^{\rm L}_{-,n}{\cal P}^{\rm R}_{-,n}{\cal U}_n,\\
&{\cal A}_{3,n}={\cal P}^{\rm L}_{-,n}{\cal P}^{\rm R}_{+,n}{\cal U}_n,\;\;
{\cal A}_{4,n}={\cal P}^{\rm L}_{+,n}{\cal P}^{\rm R}_{-,n}{\cal U}_n.
\end{split}
\end{align}
The ${\cal A}$ matrices take the form
\begin{equation}
{\cal A}_{s_n,n}= |n\rangle\langle n|\otimes M_{s_n,n}+c_{s_n,n} (1-|n\rangle\langle n|)\otimes\sigma_0,
\end{equation}
with $2\times 2$ matrices $M$ and coefficients $c$ given by
\begin{equation}
\begin{split}
&M_{1,n}=\tfrac{1}{2}\sqrt{1-w_n^2}\,U_n=M_{2,n},\\
&M_{3,n}=\tfrac{1}{2}(\sigma_0+{w_n}\sigma_z)U_n,\\
&M_{4,n}=\tfrac{1}{2}(\sigma_0-{w_n}\sigma_z)U_n,\\
&c_{1,n}=\tfrac{1}{2}(1-{w_n}),\;\;c_{2,n}=\tfrac{1}{2}(1+{w_n}),\\
&c_{3,n}=\tfrac{1}{2}\sqrt{1-w_n^2}=c_{4,n}.
\end{split}\label{Mcdef}
\end{equation}
One may check the normalization,
\begin{align}
\sum_{s=1}^4  M_{s,n}^{\dagger} M_{s,n}^{\vphantom{\dagger}} =1={}&\sum_{s=1}^4|c_{s,n}|^2\nonumber\\
&\Rightarrow\sum_{s=1}^4 {\cal A}_{s,n}^{\dagger}{\cal A}_{s,n}^{\vphantom{\dagger}}=1.
\end{align}

\subsection{From quantum trajectories to master equation}

The set of Kraus matrices ${\cal K}_{\bm s}$ implements a recursion relation for $2L\times 2L$ matrices ${\cal G}(t)$, which are updated according to
\begin{equation}
{\cal G}(t+1)=\sum_{\bm s}{\cal K}_{\bm{s}}{\cal G}(t){\cal K}_{\bm{s}}^\dagger.\label{GKrecursion}
\end{equation}
If we take as initial condition ${\cal G}(0)=|\Psi_0\rangle\langle \Psi_0|$, the transmission probability follows from
\begin{align}
{\cal T}={}&\lim_{t\rightarrow\infty}\sum_{n=L+1}^\infty \langle n+|{\cal G}(t)|n+\rangle\nonumber\\
={}&\sum_{t=0}^\infty\langle L+1,+|{\cal G}(t)|L+1,+\rangle.
\label{calTcalG}
\end{align}
In the second equality we used that sites $n>L$ have free propagation (right-moving for $\sigma=+$).

The sum over ${\bm s}$ in Eq.\ \eqref{GKrecursion} contains an exponentially large number of $4^L$ terms, each of which describes one time step of a quantum trajectory. It is possible to reduce this to order $L$ terms by purely algebraic means, see App.\ \ref{app_algebra}. The result is the master equation of a dephasing channel,
\begin{widetext}
\begin{subequations}
\label{mastereq}
\begin{align}
&{\cal G}(t+1)=e^{i\hat{k}\sigma_z}\biggl(\Omega{\cal G}(t) \Omega^\dagger +\sum_{n}w_n^2 |n\rangle\langle n|\otimes{\cal D}[U_n{\cal G}_{nn}(t)U_n^\dagger]\biggr)e^{-i\hat{k}\sigma_z},\\
&\Omega=\sum_n \sqrt{1-w_n^2} \,|n\rangle\langle n|\otimes U_n,\;\;{\cal G}_{nn}=\langle n|{\cal G}|n\rangle,\;{\cal D}[M]=\tfrac{1}{2}M+\tfrac{1}{2}\sigma_z M\sigma_z.
\end{align}
\end{subequations}
The operator ${\cal D}[M]$ introduces decoherence by zeroing out the off-diagonal elements of the pseudospin matrix $M$, leaving the diagonal elements unaffected.
\end{widetext}

\section{Solution of the master equation}
\label{sec_solution}

\subsection{Vectorization}

The master equation \eqref{mastereq} may be solved, at least formally, by vectorization of the operators --- an $L\times L$ matrix ${\cal G}$ is converted into a vector $\operatorname{vec}({\cal G})$ of length $L^2$ by stacking the columns of the matrix on top of one another. The operation $\operatorname{unvec}(\cdot)$ inverts that process.

The identities
\begin{equation}
\begin{split}
&\operatorname{vec}(ABC) = (C^\mathrm{T}\otimes A) \operatorname{vec}(B),\\
&(A_1A_2)\otimes (B_1 B_2)=(A_1\otimes B_1)(A_2\otimes B_2),\\
&\operatorname{Tr} A=(\operatorname{vec}I)^\top \operatorname{vec}A,
\end{split}
\end{equation}
allow to rewrite Eq.\ \eqref{mastereq} in the form of a matrix-vector product,
\begin{align}
&\operatorname{vec}{\cal G}(t+1)={\cal L}\operatorname{vec}{\cal G}(t),\\
&{\cal L}=\bigl( e^{i\hat{k}\sigma_z}\otimes e^{i\hat{k}\sigma_z}\bigr)\biggl[ \Omega^\ast\otimes\Omega\nonumber\\
&+\tfrac{1}{2}\sum_n w_n^2\bigl(|n\rangle\langle n|\otimes U_n^\ast)\otimes(|n\rangle\langle n|\otimes U_n\bigr)\nonumber\\
&+\tfrac{1}{2}\sum_n w_n^2\bigl(|n\rangle\langle n|\otimes\sigma_z U_n^\ast)\otimes(|n\rangle\langle n|\otimes\sigma_z U_n)\biggr],
\end{align}
with trace preserving condition
\begin{equation}
(\operatorname{vec}I)^\top {\cal L}=(\operatorname{vec}I)^\top
\end{equation}
and solution
\begin{equation}
\operatorname{vec}{\cal G}(t)={\cal L}^t\operatorname{vec}{\cal G}(0).
\end{equation}

The transmission probability then follows from Eq.\ \eqref{calTcalG},
\begin{align}
{\cal T}={}&\sum_{t=0}^\infty\langle L+1,+|\operatorname{unvec}[{\cal L}^t\operatorname{vec}{\cal G}(0)]|L+1,+\rangle\nonumber\\
={}&\langle L+1,+\bigl|\operatorname{unvec}\bigl[(1-{\cal L})^{-1}\operatorname{vec}{\cal G}(0)\bigr]\bigr|L+1,+\rangle.
\end{align}

\subsection{Elimination of the leads}

The sites in the leads, $n<1$ and $n>L$, expand the dimension of the matrix ${\cal L}$, complicating the matrix algebra. We wish to eliminate these degrees of freedom. For the right lead, $n>L$, this elimination has already been accomplished by Eq.\ \eqref{calTcalG}. Only the site $n=L+1$ remains in the formula for the transmission probability. The sites in the left lead enter via the initial state ${\cal G}(0)=|\Psi_0\rangle\langle \Psi_0|$, with $|\Psi_0\rangle$ defined in Eq.\ \eqref{Psi0def} as a sum over $L_0+1$ sites $n\leq 0$. We may eliminate these by the following algebraic procedure.

The sum
\begin{equation}
{\cal G}(0)=(1+L_0)^{-1}\sum_{n,m=-L_0}^{0}|n,+\rangle\langle m,+|
\end{equation}
corresponds to
\begin{equation}
{\cal T}=(1+L_0)^{-1}\sum_{n,m=-L_0}^0 {\cal T}_{nm},
\end{equation}
where ${\cal T}_{nm}$ is the contribution to the transmission probability ${\cal T}$ from the term $|n,+\rangle\langle m,+|$ in the initial condition ${\cal G}(0)$. 

Because of translational invariance in the leads ${\cal T}_{nm}={\cal T}_{\delta n}$ depends only on the difference $\delta n=n-m$ of the site indices $n,m\in\{-L_0,-L_0+1,\ldots ,-1,0\}$. We thus have
\begin{equation}
{\cal T}=\frac{1+L_0-|\delta n|}{1+L_0}\sum_{\delta n=-L_0}^{L_0}{\cal T}_{\delta n}.
\end{equation}
The range of $|\delta n|$ that contributes effectively to ${\cal T}$ is set by $L$. In the limit $L_0\rightarrow\infty$ at fixed $L$ we thus obtain
\begin{equation}
\lim_{L_0\rightarrow\infty}{\cal T}=\sum_{\delta n=-\infty}^{\infty}{\cal T}_{\delta n}={\cal T}_0+2\operatorname{Re}\sum_{\delta n=1}^\infty{\cal T}_{\delta n}.\label{calTdeltan3}
\end{equation}
In the second equality we used that ${\cal T}_{-\delta n}={\cal T}_{\delta n}^\ast$.

To proceed we note that the state $\rho_m(0)=|0,+\rangle\langle-m,+|$ with $m\geq 1$ evolves during the first $m$ time steps as
\begin{equation}
\begin{split}
&\rho_m(t)={\cal S}^t|0,+\rangle\langle t-m,+|,\;\;t\in\{0,1,\ldots m\},\\
&{\cal S}=e^{i\hat{k}\sigma_z}\Omega.
\end{split}
\end{equation}
The decoherence terms in the master equation \eqref{mastereq} do not contribute because $\rho_m(t)|n\rangle=0$ for all $n\geq 1$ and $t\leq m$. For the same reason $\rho_m(t)$ does not contribute to the transmission probability for times $t\leq m$. 

To compute ${\cal T}_{\delta n}$ with $\delta n\geq 1$ we may therefore replace the initial state $|0,+\rangle\langle-\delta n,0|$ by 
\begin{equation}
\rho_{\delta n}(\delta n)={\cal S}^{\delta n}\rho_0,\;\;\rho_0=|0,+\rangle \langle 0,+|.
\end{equation}
We sum the geometric series over $\delta n$,
\begin{align}
\sum_{\delta n=1}^\infty{\cal T}_{\delta n}={}&\langle L+1,+\bigl|\operatorname{unvec}(1-{\cal L})^{-1}\nonumber\\
&\operatorname{vec}{\cal S}(1-{\cal S})^{-1}\rho_0\bigr|L+1,+\rangle.\label{calTdeltan2}
\end{align}
Substitution into Eq.\ \eqref{calTdeltan3} then gives
\begin{align}
{\cal T}={}&\operatorname{Re}\langle L+1,+\bigl|\operatorname{unvec}(1-{\cal L})^{-1}\operatorname{vec}\Xi\rho_0\bigr|L+1,+\rangle,\label{calTdeltan4}\\
\Xi={}&1+2{\cal S}(1-{\cal S})^{-1}=(1+{\cal S})(1-{\cal S})^{-1}.
\end{align}
This expression for the transmission probability only involves sites $n=0,1,2\ldots L,L+1$, eliminating the infinite leads.

For later use we note that, instead of taking the real part, we can work with the full expression if we define the ``lead operator''
\begin{equation}
{\cal L}_{\rm lead}[\rho]=\tfrac{1}{2}\Xi\rho+\tfrac{1}{2}\rho \Xi^\dagger,\label{leadoperator}
\end{equation}
when Eq.\ \eqref{calTdeltan4} can be rewritten equivalently as
\begin{align}
{\cal T}={}&\langle L+1,+\bigl|\operatorname{unvec}(1-{\cal L})^{-1}\nonumber\\
&\operatorname{vec}{\cal L}_{\rm lead}[\rho_0]\bigr| L+1,+\rangle.\label{calTdeltan5}
\end{align}

\section{Tunnel barriers in series}
\label{sec_tunnel}

Closed form expressions can be obtained for a small number of tunnel barriers in series. We study that first, before considering the numerics of a disordered conductor. 

\subsection{Resonant tunneling}

We consider $N$ tunnel barriers in series, each with scattering matrix $U_n=\sigma_x S_n$ parameterized by
\begin{align}
S_n={}&\begin{pmatrix}
e^{i\phi_{n}}&0\\
0&e^{i\phi'_{n}}
\end{pmatrix}\begin{pmatrix}
\sqrt{1-\Gamma_n}&\sqrt{\Gamma_n}\\
\sqrt{\Gamma_n}&-\sqrt{1-\Gamma_n}
\end{pmatrix}\nonumber\\
&\cdot\begin{pmatrix}
e^{i\psi_{n}}&0\\
0&e^{i\psi'_{n}}
\end{pmatrix}.\label{Sndef}
\end{align}
The tunnel probability of the $n$-th barrier is $\Gamma_n\in(0,1)$. The phase shifts accumulated upon propagation from one barrier to the next are included in $S_n$ via the four phase factors.

The unitaries $U_n$ and $S_n$ are related by a change of basis, such that for $\Gamma_n=1$, so for a transparent barrier, $U_n$ is fully diagonal while $S_n$ is fully off-diagonal. Time reversal symmetry is expressed differently in the two bases,
\begin{equation}
S_n^\top=S_n,\;\;U_n^\top=\sigma_x U_n\sigma_x,\label{TRS}
\end{equation}
implying $\phi_n=\psi_n$ and $\phi'_n=\psi'_n$ in the presence of time reversal symmetry. In the one-dimensional setting considered here the transmission probability does not depend on whether time reversal symmetry is broken or not.

The vectorized master equation \eqref{calTdeltan4} can be solved efficiently because the matrix ${\cal L}$ is sparse. For $N=2$ we find 
\begin{equation}
{\cal T}=\frac{\Gamma_1\Gamma_2}{1+\gamma^2\rho_1\rho_2+2 \gamma\sqrt{\rho_1\rho_2}\cos\phi}\,\frac{1-\gamma^2\rho_1\rho_2}{1-\rho_1\rho_2},\label{TNis2}
\end{equation}
with $\phi=\phi'_1+\psi'_1+\phi_2+\psi_2$, $\rho_n=1-\Gamma_n$, $\gamma=1-w^2$. This interpolates between the phase coherent resonant tunneling transmission
\begin{equation}
{\cal T}_{\rm coh}=\left|\frac{\sqrt{\Gamma_1\Gamma_2}}{1+\sqrt{\rho_1\rho_2}e^{i\phi}}\right|^2=
\frac{\Gamma_1\Gamma_2}{1+\rho_1\rho_2+2\sqrt{\rho_1\rho_2}\cos\phi}
\end{equation}
for $w=0$ and the incoherent series transmission
\begin{equation}
{\cal T}_{\rm incoh}=\left(1+\sum_{n}(1/\Gamma_n-1)\right)^{-1}=\frac{\Gamma_1\Gamma_2}{1-\rho_1\rho_2}\label{Tincoherent}
\end{equation}
for $w=1$. The $N=2$ result \eqref{TNis2} is identical to that following from the voltage probe model, see App.\ \ref{app_probe}.

\begin{figure}[tb]
\centerline{\includegraphics[width=0.9\linewidth]{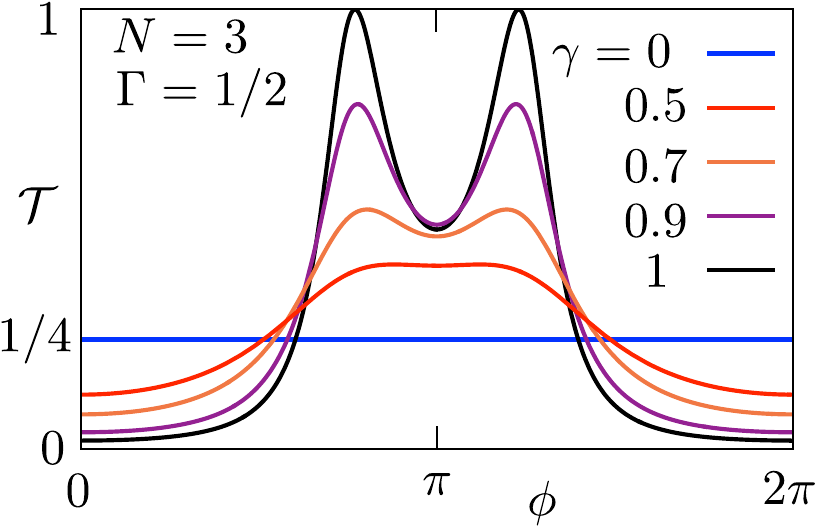}}
\caption{Scattering phase shift $\phi$ dependence of the transmission probability ${\cal T}$ through $N=3$ identical tunnel barriers in series, each with tunnel probability $\Gamma=1-\rho=1/2$. The effect of decoherence with strength $w=\sqrt{1-\gamma}$ is computed from Eq.\ \eqref{TNis3}.
}
\label{fig_barriers}
\end{figure}

For $N=3$ we simplify the formula by taking identical tunnel barriers, $\Gamma_n=\Gamma=1-\rho$, $\phi_n+\phi'_n+\psi_n+\psi'_n=\phi$, for $n\in\{1,2,3\}$. We find
\begin{widetext}
\begin{equation}
{\cal T}=\frac{1-{\rho}}{1+{\gamma}^2 {\rho} +2 {\rho} }\frac{ 1+\gamma^2\rho+ {\gamma}^3 {\rho}^3(4-{\gamma}^3) -{\gamma}^3 {\rho}^2({\gamma}+4)+2{\gamma} {\rho}  (1-{\gamma}) ({\gamma} {\rho}+1) ({\gamma}^2 {\rho}+1) \cos {\phi}}{1+{\gamma}^2 {\rho}^2({\gamma}^2+4) + 4\gamma\rho({\gamma}^2 {\rho}+1) \cos {\phi}+2{\gamma}^2\rho \cos 2 {\phi}}.\label{TNis3}
\end{equation}
\end{widetext}
In App.\ \ref{app_probe} we show that the $N=3$ formula \eqref{TNis3} is very close to, but not identical, to what would follow from the voltage probe model.

The effect of decoherence is plotted in Fig.\ \ref{fig_barriers}, for $\Gamma=1/2$. The two quasi-bound states between the first and second barrier and between the second and third barrier hybridize and split, resulting in two peaks of resonant transmission at $\phi=\pi\pm\arccos(1-\Gamma/2)\approx\pi\pm 0.72$. Dephasing broadens and lowers the peaks, they merge into a single broad maximum at $\phi=\pi$, and finally level off at the $\phi$-independent value ${\cal T}_{\rm incoh}=(1+3/\Gamma-3)^{-1}=1/4$.

\subsection{Reciprocity breaking}

\begin{figure}[tb]
\centerline{\includegraphics[width=0.8\linewidth]{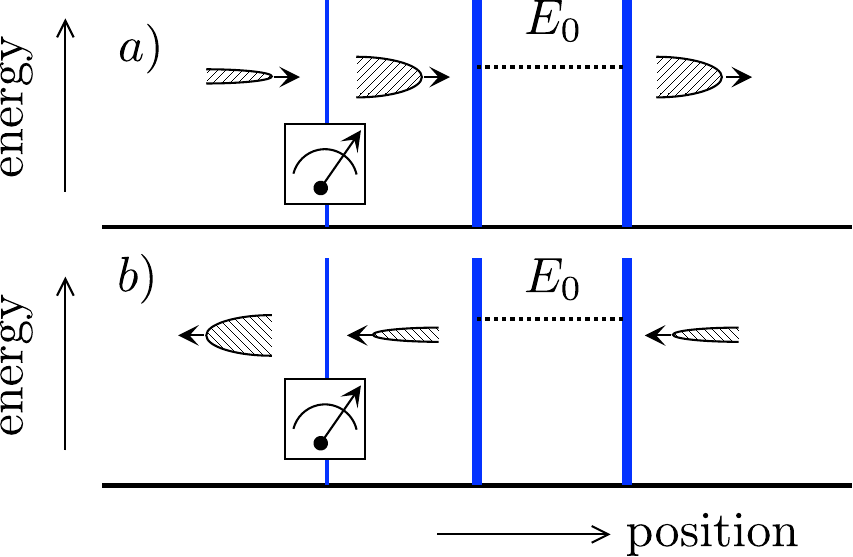}}
\caption{Illustration of reciprocity breaking by monitoring \cite{Fer24}. Shown is a three-barrier structure (blue) without left-right symmetry (a thin barrier at the left, two thick barriers at the right). Time-resolved monitoring (on a time scale $\tau_c$) measures the outcome of the scattering by the thin barrier, broadening the wave packet of the transmitted electron (from $eV$ to $\hbar/\tau_c$). The broadened wave packet overlaps with the bound state $E_0$ in between the thick barriers. This increases the transmission probability if monitoring happens \textit{before} the electron reaches the thick barriers (panel a), but not if it happens \textit{after} (panel b).
}
\label{fig_reciprocity_diagram}
\end{figure}

As explained by Ferreira \textit{et al.} \cite{Fer24}, the time-resolved monitoring of the occupation number of a quasi-bound state is an inelastic process that drives the conductor out of equilibrium, even in the absence of a voltage bias. Reciprocity may be broken: an electron injected from the left reservoir may have a different transmission probability than an electron injected from the right reservoir. Fig.\ \ref{fig_reciprocity_diagram} illustrates the mechanism.

The $N=2$ formula \eqref{TNis2} satisfies reciprocity, it is invariant if $\Gamma_1$ and $\Gamma_2$ are interchanged. To check for reciprocity when $N=3$ we take different tunnel probabilities $\Gamma_n=1-\rho_n$. A relatively compact expression results if we set all phase shifts to zero, so real $S_n$:
\begin{widetext}
\begin{equation}
{\cal T}=\frac{\Gamma_1\Gamma_2\Gamma_3 \bigl(1-\gamma^4 \rho_1 \rho_3-\gamma \sqrt{\rho_2} \biglb[2 \gamma (1-\rho_1) \sqrt{\rho_3}+(\sqrt{\rho_1}-\sqrt{\rho_3}) (\gamma^2 \sqrt{\rho_1\rho_3}+1)\bigrb]\bigr)}{\bigl(\gamma^2 \sqrt{\rho_1\rho_3}+\gamma \sqrt{\rho_2} (\sqrt{\rho_1}+\sqrt{\rho_3})+1\bigr) \bigl((1-\rho_1 \rho_3) (\gamma^2 \sqrt{\rho_1\rho_3}+1)+\rho_2 (\gamma^2 \sqrt{\rho_1\rho_3} (\rho_1+\rho_3-2)+2 \rho_1 \rho_3-\rho_1-\rho_3)\bigr)}.
\end{equation}
\end{widetext}
The term in square brackets in the numerator is not symmetric upon exchange of $\rho_1$ and $\rho_3$, so reciprocity is broken. In particular, in the limit $\rho_1\rightarrow 0$ of a fully transmitting first barrier one has
\begin{equation}
{\cal T}\bigr|_{\rho_1=0}=\frac{\Gamma_2\Gamma_3\biglb[1-\gamma\sqrt{\rho_2\rho_3}+2\gamma(1-\gamma)\sqrt{\rho_2\rho_3}\bigrb]}{(1-\rho_2\rho_3)(1+\gamma\sqrt{\rho_2\rho_3})},
\end{equation}
while the opposite case $\rho_3\rightarrow 0$ of a fully transmitting last barrier gives a smaller transmission probability,
\begin{equation}
{\cal T}\bigr|_{\rho_3=0}=\frac{\Gamma_1\Gamma_2\biglb[1-\gamma\sqrt{\rho_1\rho_2}\bigrb]}{(1-\rho_1\rho_2)(1+\gamma\sqrt{\rho_1\rho_2})},
\end{equation}
provided that $\gamma$ differs from 0 or 1.

The reciprocity breaking implies that the differential conductance $G=dI/dV$ depends on how the voltage difference $V=V_{\rm L}-V_{\rm R}$ between the left and right contacts is divided. We focus on the case that the right contact is grounded, $V_{\rm R}=0$, and the voltage bias $V=V_{\rm L}$ is applied to the left contact. Eq.\ \eqref{GPdef} refers to that case.

\section{Localization by disorder}
\label{sec_localization}

\subsection{Full phase coherence}

We work in the weak disorder regime, when 
\begin{equation}
\ell/\delta L=\frac{\Gamma_n}{1-\Gamma_n}\gg 1,\label{elldef}
\end{equation}
with $\ell$ the mean free path \cite{note4} and $\delta L$ the length of the $n$-th segment. The entire conductor has length $L=N\delta L$. Disorder is introduced in the scattering matrix parameterization \eqref{Sndef} by drawing the phases $\phi_{n,1},\phi_{n,2},\phi_{n,3},\phi_{n,4}$ uniformly and independently from the interval $(0,2\pi)$. The tunnel probabilities are all taken to be the same $\Gamma_n=\Gamma$.

In case of full phase coherence one has \cite{Ger59,Abr81,Bee97}
\begin{equation}
\mathbb{E}[\ln{\cal T}]=-L/\ell,\label{xidef}
\end{equation}
which defines the localization length $\xi=\ell$. All states are localized, the diffusive regime $\ell\ll L\ll\xi$ does not exist in 1D.

The full distribution of ${\cal T}$ is broad (log-normal), so that the transmission probability which is typical for a single sample is not the mean $\mathbb{E}[{\cal T}]$ but the logarithmic mean
\begin{equation}
{\cal T}_{\rm typ}=e^{\mathbb{E}[\ln{\cal T}]}=e^{-L/\xi},\;\;\xi=\ell.
\end{equation}

\subsection{Coherence length}

The master equation \eqref{mastereq} alternates coherent evolution with projective measurements, happening independently with probability $w^2$ at each time step $\delta t=\delta L/v_{\rm F}$ (where $v_{\rm F}$ is the Fermi velocity). The coherence time
\begin{equation}
\tau_\phi=\delta t/w^2
\end{equation}
is the mean time between measurements. We wish to relate that to a coherence length $\ell_\phi$, the typical length of a phase coherent segment of the conductor. In a diffusive conductor one would have $\ell_\phi=\sqrt{D\tau_\phi}$, with $D$ the diffusion constant. This relationship needs modification in a 1D conductor, without a diffusive regime. We argue as follows.

The Thouless formula relates the conductance $G\simeq (e^2/h)E_{\rm T}/\delta E$ to the level spacing $\delta E$ and the Thouless energy $E_{\rm T}=\hbar/\tau_{\rm dwell}$, defined in terms of the time $\tau_{\rm dwell}$ that an electron spends inside the conductor \cite{Imry}. Setting $G\simeq (e^2/h){\cal T}_{\rm typ}=(e^2/h)e^{-L/\xi}$, $\delta E\simeq \hbar v_{\rm F}/L$, we relate $\tau_{\rm dwell}\mapsto\tau_\phi$ to $L\mapsto\ell_\phi$ via
\begin{equation}
e^{-\ell_\phi/\xi}=\frac{\ell_\phi}{v_{\rm F}\tau_\phi}\Rightarrow \ell_\phi=\xi W_{0}(v_{\rm F}\tau_\phi/\xi),\label{W0result}
\end{equation}
where $W_{0}(x)$ is the Lambert W-function. Asymptotically for large $\tau_\phi$ this function grows logarithmically,
\begin{equation}
\ell_\phi\underset{\tau_\phi\rightarrow\infty}{\longrightarrow} \xi\ln(v_{\rm F}\tau_\phi/\xi) \label{lphi}.
\end{equation}

\begin{figure}[tb]
\centerline{\includegraphics[width=0.7\linewidth]{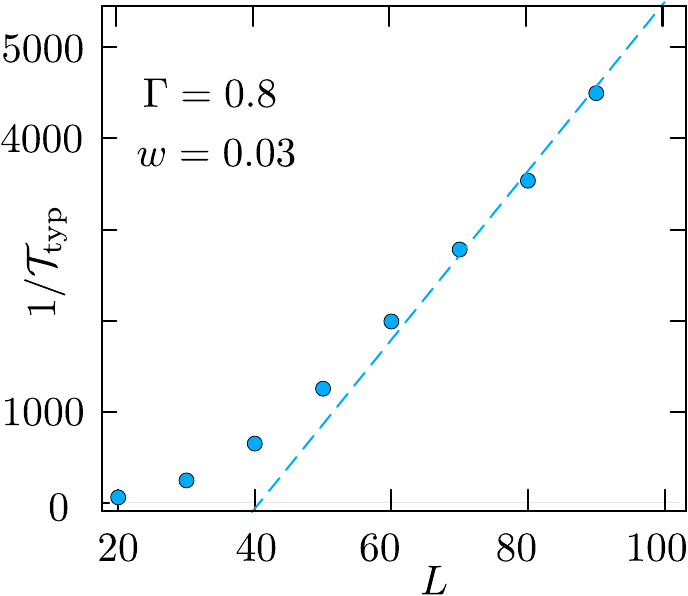}}
\caption{Reciprocal of the typical transmission probability ${\cal T}_{\rm typ}=e^{\mathbb{E}[\ln {\cal T}]}$ as a function of conductor length $L$. The data points have been computed from Eq.\ \eqref{calTdeltan5} for tunnel probability $\Gamma=0.8$ and measurement strength $w=0.03$. The expectation value $\mathbb{E}[\ln {\cal T}]$ is an average over $10^4$ random disorder realizations. The dashed line is a linear fit to the large-$L$ data, to show the crossover to the diffusive regime.}
\label{fig_RvsL}
\end{figure}

\begin{figure}[tb]
\centerline{\includegraphics[width=0.7\linewidth]{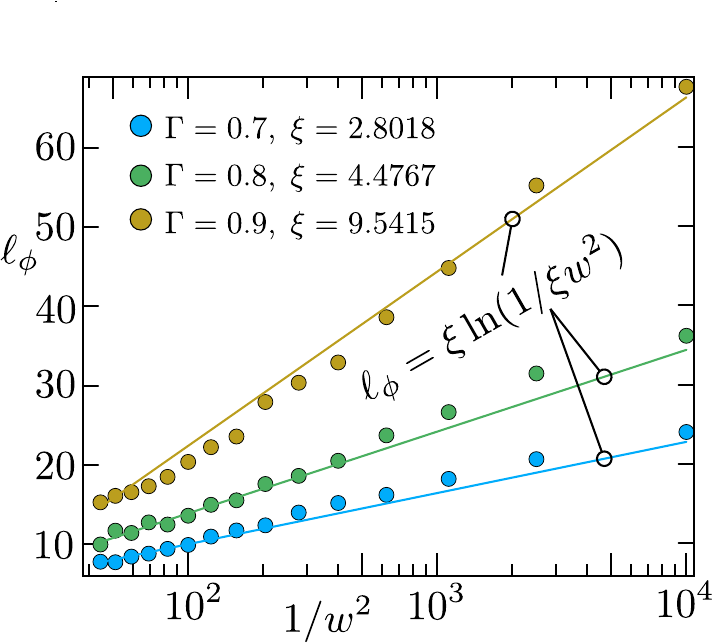}}
\caption{Coherence length $\ell_\phi$ as a function of $1/w^2$ for three values of the tunnel probability. The solid lines are given by Eq.\ \eqref{lphi}, which in dimensionless units reads $\ell_\phi=\xi\ln(1/\xi w^2)$. The localization length $\xi$ is not fitted to this data, it is obtained independently from the exponential decay of ${\cal T}_{\rm typ}\propto e^{-L/\xi}$ for $w=0$.}
\label{fig_lphivsw}
\end{figure}

\subsection{Numerical results}

Our numerical results are shown in Figs.\ \ref{fig_RvsL} and \ref{fig_lphivsw}. We fit the small-$L$ data for $1/{\cal T}_{\rm typ}$ to an exponential function. We then perform a large-$L$ linear fit of $1/{\cal T}_{\rm typ}$ (shown in Fig.\ \ref{fig_RvsL}). We estimate the coherence length $\ell_{\phi}$ from the value of $L$ where the small-$L$ and the large-$L$ fitting functions have the smallest difference. The resulting $w$-dependence of $\ell_\phi$ is close to logarithmic, see Fig.\ \ref{fig_lphivsw}.

The close agreement is somewhat surprising, in view of the qualitative nature of the argument leading to Eq.\ \eqref{lphi}. There is some uncertainty in the way we define the coherence length $\ell_\phi$, since the crossover region from exponential to linear $L$-dependence is rather broad. We have tried different ways to extract $\ell_\phi$ from the data and consistently find a linear dependence of $\ell_\phi$ on $\ln\tau_\phi$, which we believe is a trustworthy result in the $\tau_\phi\rightarrow\infty$ limit.

\section{Conclusion}
\label{sec_conclude}

In conclusion, we have removed the restriction to chiral motion in the monitored transport formalism of Ref.\ \onlinecite{Bee25}, and applied the theory to the problem of 1D localization by disorder. Time-resolved measurements of the scattering outcome define a quantum channel (a completely-positive trace-preserving map). Although the number of measurement outcomes grows exponentially with system size, their contribution to the single-particle transmission probability can be resummed algebraically into the master equation \eqref{mastereq}.

In the zero-temperature, linear-response regime the full counting statistics of transmitted charge remains binomial: Monitoring modifies the transmission probability ${\cal T}$, but not the functional form of the counting distribution. The central effect of monitoring is to cut off the coherent interference responsible for one-dimensional localization. The typical transmission probability crosses over from ${\cal T}_{\rm typ}\propto e^{-L/\xi}$ to a diffusive $1/L$ tail. Our numerical solution of the quantum master equation indicates a logarithmic dependence of $\ell_\phi$ on the monitoring strength, in accord with a qualitative argument.

An analytical solution of the problem remains open. The usual scaling approach \cite{Ger59,Abr81} relies essentially on unitary dynamics, which does not carry over to the monitored quantum channel. It would be interesting to adapt Keldysh methods \cite{Jin22} to this setting, and also to lift the algebraic resummation of quantum trajectories from the single-particle sector to the full many-particle Fock space.

\acknowledgments

We have benefited from discussions with T. Giamarchi.\\
This work was supported by the Netherlands Organisation for Scientific Research (NWO/OCW), as part of Quantum Limits (project number {\sc summit}.1.1016). 

\appendix

\section{Kraus matrix algebra}
\label{app_algebra}

We work out the algebra that transforms the sum \eqref{GKrecursion} over quantum trajectories into the master equation \eqref{mastereq}.

The Kraus matrices act on $\mathbb{C}^L\otimes \mathbb{C}^2$, representing a lattice of $L$ sites with a pseudospin degree of freedom. Each Kraus matrix \cite{note2}
\begin{equation}
{\cal K}_{\bm s}=\prod_{n=1}^L {\cal A}_{s_n,n}\label{KsAsnn}
\end{equation}
is labeled by a string $\bm{s}=\{s_1,s_2,\ldots s_L\}$, where the index $s_n$ selects a matrix ${\cal A}_{s_n,n}$ of the form
\begin{equation}
{\cal A}_{s_n,n}= P_n\otimes M_{s_n,n}+c_{s_n,n} Q_n\otimes\sigma_0.\label{Asnn}
\end{equation}
Here $P_n=|n\rangle\langle n|=1-Q_n$ is the projector on site $n$, $c_{s_n,n}$ is a complex coefficient, and $M_{s_n,n}$ is a $2\times 2$ complex matrix (not necessarily unitary). The normalization
\begin{equation}
\sum_{s_n}  M_{s_n,n}^{\dagger} M_{s_n,n}^{\vphantom{\dagger}} =\sigma_0,\;\;\sum_{s_n}|c_{s_n,n}|^2=1,\label{sumcndn}
\end{equation}
ensures the sum rule
\begin{equation}
\sum_{\bm s}{\cal K}_{\bm s}^\dagger {\cal K}^{\vphantom{\dagger}}_{\bm s}=1.
\end{equation}

Substitution of Eq.\ \eqref{Asnn} into the Kraus matrix \eqref{KsAsnn} produces a sum of products of $P$ and $Q$ projectors. Each product cannot contain more than one $P$ projector, because the projectors commute and $P_n P_m=0$ if $n\neq m$. Moreover, $\prod_{n=1}^L Q_n =0$, so each product must contain precisely one $P$ projector, say $P_n$. Summation over the products then gives
\begin{align}
K_{\bm s}={}& \sum_{n=1}^L (P_n\otimes M_{s_n,n})\prod_{m\neq n}c_{s_m,m}(Q_m\otimes \sigma_0)\nonumber\\
={}&\sum_{n=1}^L (P_n\otimes M_{s_n,n})\prod_{m\neq n}c_{s_m,m},
\end{align}
where in the second equality we used that 
\begin{align}
&(P_n\otimes M_{s_n,n})(Q_m\otimes \sigma_0)=(P_nQ_m)\otimes(M_{s_n,n}\sigma_0)\nonumber\\
&\qquad=P_n\otimes M_{s_n,n},\;\;\text{for}\;\;n\neq m.
\end{align}
We thus have
\begin{widetext}
\begin{equation}
\sum_{\bm s}{\cal K}^{\vphantom{\dagger}}_{\bm s}{\cal G} {\cal K}_{\bm s}^\dagger=\sum_{\bm s}\left(\sum_{n=1}^L (P_n\otimes M_{s_n,n})\prod_{i\neq n}c_{s_i,i}\right){\cal G}\left(\sum_{m=1}^L (P_m\otimes M_{s_m,m})\prod_{j\neq m}c_{s_j,j}\right)^\dagger.
\end{equation}

The sum over the Kraus matrix indices $\bm{s}$ can be worked out further by splitting the double sum over the lattice sites into $n=m$ and $n\neq m$,
\begin{align}
\sum_{\bm s}{\cal K}^{\vphantom{\dagger}}_{\bm s}{\cal G} {\cal K}_{\bm s}^\dagger={}&\sum_{n\neq m}\sum_{s_n,s_m}\bigl(c_{s_m,m}(P_n\otimes M_{s_n,n})\bigr){\cal G}\bigl( c_{s_n,n}(P_m\otimes M_{s_m,m})\bigr)^\dagger
+\sum_{n} \sum_{s_n} (P_n\otimes M_{s_n,n}){\cal G}(P_n\otimes M_{s_n,n})^\dagger\nonumber\\
={}&\sum_{n\neq m}(P_n\otimes\Omega_n^{\vphantom{\dagger}}){\cal G}(P_m\otimes\Omega_m)^\dagger+\sum_{n}\sum_{s_n}(P_n\otimes M_{s_n,n}){\cal G}(P_n\otimes M_{s_n,n})^\dagger,\;\;
\Omega_n=\sum_{s_n} c_{s_n,n}^\ast M_{s_n,n},
\end{align}
where we have used the normalization \eqref{sumcndn}. One more step, to remove the restriction $n\neq m$,
\begin{equation}
\sum_{\bm s}{\cal K}^{\vphantom{\dagger}}_{\bm s}{\cal G} {\cal K}_{\bm s}^\dagger=\sum_{n,m}|n\rangle\langle m|\otimes\Omega_n^{\vphantom{\dagger}}{\cal G}_{nm}\Omega_m^\dagger+\sum_{n}|n\rangle\langle n|\otimes\left(\sum_{s_n}  M_{s_n,n}^{\vphantom{\dagger}}{\cal G}_{nn} M_{s_n,n}^\dagger-\Omega_n{\cal G}_{nn}\Omega_n^\dagger\right),
\end{equation}
with ${\cal G}_{nm}=\langle n|{\cal G}|m\rangle$.
\end{widetext}

We then substitute the four matrices $M$ and four coefficients $c$ from Eq.\ \eqref{Mcdef}, which give
\begin{align}
&\Omega_n=\sqrt{1-w_n^2}\,U_n,\\
&\sum_{s_n}  M_{s_n,n}^{\vphantom{\dagger}}{\cal G}_{nn} M_{s_n,n}^\dagger-\Omega_n{\cal G}_{nn}\Omega_n^\dagger\nonumber\\
&\qquad=\tfrac{1}{2}w_n^2 U_n{\cal G}_{nn}U_n^\dagger + \tfrac{1}{2}w_n^2 \sigma_z U_n{\cal G}_{nn}U_n^\dagger\sigma_z.
\end{align}
This is the master equation \eqref{mastereq} from the main text.

\section{Comparison with voltage probe model}
\label{app_probe}

The model of decoherence introduced by B\"{u}ttiker inserts additional leads (voltage probes) in between scattering centers. Each lead is connected to an electron reservoir in thermal equilibrium at a voltage which is adjusted so that the net current through the lead vanishes. We compare results from that model to those obtained from the master equation. 

\begin{figure}[tb]
\centerline{\includegraphics[width=0.8\linewidth]{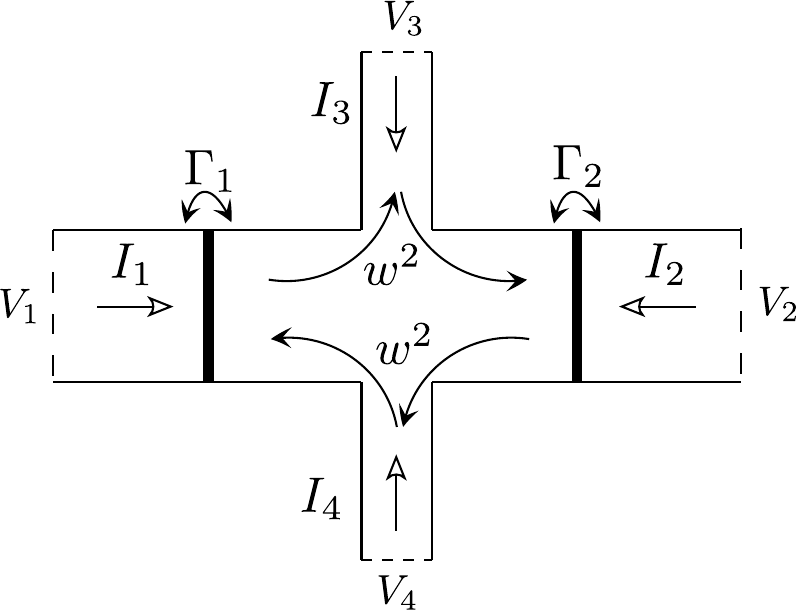}}
\caption{Illustration of the voltage probe model of dephasing, for the case of two tunnel barriers (tunnel probability $\Gamma_n=1-\rho_n$). The barriers are in series in a four-terminal geometry. Terminals 1 and 2 are the current source and drain. Terminals 3 and 4 are voltage probes, which serve to introduce dephasing without backscattering. A right-moving electron enters probe 3 with probability $w^2=1-\gamma$, a left-moving electron enters probe 4 with the same probability. Voltages $V_3$ and $V_4$ are adjusted so that $I_3=0=I_4$. If more tunnel barriers are added in series, an additional pair of voltage probes is introduced between subsequent barriers.
}
\label{fig_double_barrier}
\end{figure}

We consider $N$ tunnel barriers in series, each with scattering matrix $S_n$ given by Eq. \eqref{Sndef}. To model dephasing without backscattering we introduce two separate voltage probes in between subsequent barriers, one which couples exclusively to left-movers, the other to right-movers (see Fig.\ \ref{fig_double_barrier}). Electrons enter each voltage probe with the same probability $w^2=1-\gamma$. The four-terminal scattering matrix of the region in between two barriers, with the leads numbered as in Fig.\ \ref{fig_double_barrier}, is
\begin{equation}
S_{w}=\begin{pmatrix}
0&\sqrt{1-w^2}&0&-w\\
\sqrt{1-w^2}&0&-w&0\\
w&0&\sqrt{1-w^2}&0\\
0&w&0&\sqrt{1-w^2}
\end{pmatrix}.
\end{equation}
It is not symmetric, the construction necessarily breaks time reversal symmetry.

The transmission probabilities $T_{ij}$ from reservoir $j$ to reservoir $i$ are computed by composing the scattering matrices. The current in lead $i$ is then given  in terms of the voltages $V_j$ of the reservoirs by 
\begin{equation}
(h/e^2)I_i=(1-T_{ii})V_i-\sum_{j\neq i}T_{ij}V_j.\label{IVeqs}
\end{equation}
For $N$ barriers there are $2(N-1)$ voltage probes, including the current source and drain there are $2N$ leads. If we label the current source by index $i=1$ and the current drain by index $i=2$, the voltages $V_1=V$ and $V_{2}=0$ are given. The remaining voltages are determined from the set of equations \eqref{IVeqs} by setting $I_i=0$ for $i=3,4\ldots 2N$. Current conservation then ensures that $I_1=-I_2\equiv I$. Finally, the transmission probability ${\cal T}_{\text{probe}}$ in the voltage probe model follows from 
\begin{equation}
{\cal T}_{\text{probe}}=(h/e^2)I/V.
\end{equation}

\begin{figure}[tb]
\centerline{\includegraphics[width=0.9\linewidth]{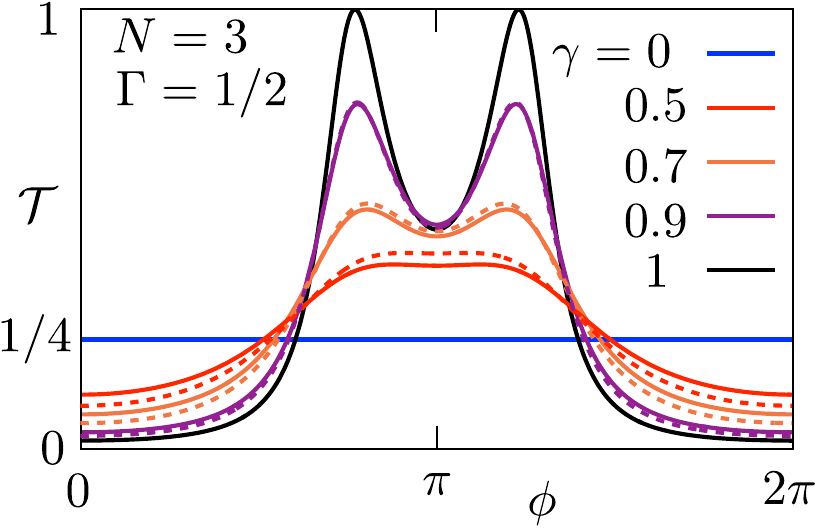}}
\caption{Same as Fig.\ \ref{fig_barriers}, comparing the result \eqref{TNis3} from the master equation (solid curves) with Eq.\ \eqref{TNis3probe} from the voltage probe model (dashed curves). The two models of decoherence differ slightly for $\gamma\neq 0,1$. 
}
\label{fig_compare}
\end{figure}

For $N=2$ we find
\begin{equation}
{\cal T}_{\text{probe}}=\frac{\Gamma_1\Gamma_2}{1+\gamma^2\rho_1\rho_2+2 \gamma\sqrt{\rho_1\rho_2}\cos\phi}\,\frac{1-\gamma^2\rho_1\rho_2}{1-\rho_1\rho_2}.
\end{equation}
The same result follows from a model with a randomly fluctuating scattering phase \cite{Zhe06}.

For $N=3$ identical barriers we find
 \begin{widetext}
\begin{equation}
{\cal T}_{\text{probe}}=\frac{(1-{\rho}) (1-{\gamma}^2 {\rho})^2}{1+(2-{\gamma}^2) {\rho}+2 {\gamma} {\rho} \cos {\phi}}\,\frac{1+{\gamma}^2 {\rho}+2 {\gamma} {\rho} \cos {\phi}}{ 1+\gamma^2\rho^2(\gamma^2+4)+4 {\gamma} {\rho} ({\gamma}^2 {\rho}+1) \cos {\phi}+2 {\gamma}^2 {\rho} \cos 2 {\phi}}.\label{TNis3probe}
\end{equation}
\end{widetext}
The $N=2$ result from the voltage probe model is the same as the result \eqref{TNis2} from the master equation, the $N=3$ result differs from the master equation formula \eqref{TNis3}. The difference is small but nonzero, see Fig.\ \ref{fig_compare}.

\section{Reciprocity preserving master equation}

The master equation \eqref{mastereq} breaks reciprocity in the weak backscattering (strong coupling) regime. We may restore that by inserting, at each measurement, the lead operator \eqref{leadoperator}. This operator ${\cal L}_{\rm lead}$ acts on a diagonal matrix, because the measurement removes all off-diagonal elements of $\rho$. To ensure that it preserves the trace we normalize \cite{note3},
\begin{equation}
\begin{split}
\tilde{\cal L}_{\rm lead}[\rho] &= \sum_{n,\sigma} \frac{\langle n,\sigma|\rho|n,\sigma\rangle}{\langle n,\sigma|\Xi+\Xi^\dagger|n,\sigma\rangle}\bigl(\Xi P_{n,\sigma}+P_{n,\sigma}\Xi^\dagger\bigr)\\
&\Rightarrow\operatorname{Tr}\tilde{\cal L}_{\rm lead}[\rho]=\operatorname{Tr}\rho,
\end{split}
\end{equation}
where we have defined the projector $P_{n,\sigma}=|n,\sigma\rangle\langle n,\sigma|$.

The modified time evolution,
\begin{widetext}
\begin{equation}
{\cal G}(t+1)=e^{i\hat{k}\sigma_z}\Omega{\cal G}(t) \Omega^\dagger e^{-i\hat{k}\sigma_z}
+\tilde{\cal L}_{\rm lead}\biggr[e^{i\hat{k}\sigma_z}\biggr(\sum_{n}w_n^2 |n\rangle\langle n|\otimes{\cal D}[U_n{\cal G}_{nn}(t)U_n^\dagger]\biggr)e^{-i\hat{k}\sigma_z}\biggr],
\end{equation}
\end{widetext}
has the form of a Redfield master equation in discrete time. It preserves Hermiticity and trace of the density matrix, but it does not preserve the positivity --- that is broken by the lead operator. Reciprocity is restored, and the results coincide precisely with the voltage probe model.


\begin{thebibliography}{99}
\bibitem{note1} In earlier related work \cite{Bee25} a different parameterization $\varepsilon$ of the decoherence strength was used, related to the parameter ${w}$ used in Eq.\ \eqref{PRLdef} by $w=\varepsilon\sqrt{2-\varepsilon^2}$. The $w$-parameterization simplifies some of our expressions.
\bibitem{note4} The transport mean free path $\ell_{\rm tr}$ of kinetic theory, defined by the average distance after which the particle's direction of motion is randomized, is related to $\ell$ by $\ell_{\rm tr}=\ell/2$, see Ref.\ \onlinecite{note4}. We also note that our definition \eqref{xidef} of the localization length differs by a factor of two from some other definitions used in the literature.
\bibitem{note2} In App.\ \ref{app_algebra} we do not include the shift operator $e^{i\hat{k}\sigma_z}$ in the definition of the Kraus matrix. This operator can be inserted at the very end of the calculation, it does not affect the algebra.
\bibitem{note3} Since $\langle 0,+|\Xi+\Xi^\dagger|0,+\rangle=1$, one has $\tilde{\cal L}_{\rm lead}[\rho_0]={\cal L}_{\rm lead}[\rho_0]$ for $\rho_0=|0,+\rangle\langle 0,+|$, and hence no normalization is needed when the lead operator is applied in Eq.\ \eqref{calTdeltan5}.
\end{thebibliography}
\end{document}